\documentclass[12pt]{article}
\newcommand{\beq}{\begin{equation}}
\newcommand{\eeq}{\end{equation}}
\newcommand{\bra}{\begin{array}}
\newcommand{\era}{\end{array}}
\newcommand{\be}{\beta}
\newcommand{\te}{\theta}
\newcommand{\al}{\alpha}
\newcommand{\ga}{\gamma}
\newcommand{\de}{\delta}
\newcommand{\da}{\dagger}

\newcommand{\om}{\omega}

\newcommand{\si}{\sigma}
\newcommand{\ep}{\epsilon}
\author{Jamila Douari \footnote{douari@sun.ac.za} \\
\small\it Stellenbosch Institute for Advanced Study, Private Bag X1,\rm\\
\small\it Matieland, Stellenbosch, 7601, South Africa\rm }
\title{Planar System and $w_\infty$ Algebra}
\begin{document}
\maketitle
\vspace*{0.5cm}

\section*{Abstract}
\hspace{.3in}We study the exotic particles symmetry in the background of noncommutative two-dimensional phase-space leading to realize in physicswise the deformed version of $C_{\lambda}$-extended Heisenberg algebra and $\om_\infty$ symmetry.

\section{Introduction}
\hspace{.3in}The terminology of $w_\infty$ algebra is used to refer to a wide class of higher spin algebras in two dimensions as a particular generalization of the Virasoro algebra \cite{pr,sm,sup}. Thus, in this work, we will look for a $w_\infty$ symmetry describing two-dimensional particles system living in non-commutative phase-space. This system consists of quasi-particles known as excitations and quasi-particles or anyons; i.e. fermions (bosons) carrying odd (even) number of elementary magnetic flux quanta. They are living in two-dimensional space as composite particles having arbitrary spin, and they are characterized by fractional statistics which are interpolating between bosonic statistics and fermionic one \cite{lm}. First of all, the present work is devoted to study the symmetry describing these special particles basing on a noncommutative geometry depending on the statistical parameter $\nu$. This latter parameter characterizes the planar system. We define the corresponding annihiliation and creation operators by introducing an operator acting in the phase direction denoted $\xi_i$. Then, the exotic particles symmetry is given having two extremes the bosonic algebra corresponding to $\nu\longrightarrow 0$ and deformed fermionic algebra for $\nu\longrightarrow 1$. The deformed $C_{\lambda}$-extended Heisenberg algebra is realized in two-dimensional noncommutative phase-space describing anyons and we construct the $\om_\infty$ symmetry on the same space basing on the "exotic" annihilation and creation operators.\\

This paper is organized as follows: In section 2, we give a brief introduction on the planar system. In section 3, we study the exotic paticles symmetry in the background of noncommutative phase-space. In section 4, we review in brief the $C_{\lambda}$-extended oscillator algebras and we give its deformed version realized on two-dimensional noncommutative phase-space and then we construct a Fock representation for the noncommutative geometry discussed in section 3. Thus, the section 4 is devoted to construct the $\om_\infty$ algebra characterizing the two-dimensional noncommutative system.

\section{Planar System}
\hspace{.3in}Let us start by recalling the origin of anyons \cite{lm}. These particles are known as excitations in two-dimensional space obey intermediate statistics that interpolate between bosonic and fermionic statistics. Thus, to see how come anyons as a theory, firstly, as known the configuration space $M^d _N$ of $N$ identical particles in $d$-dimensional space $(\Re^d )^N$  is given as follows
$$M^d _N =\frac{(\Re^d )^N -\Delta}{S_N }$$
by removing the diagonal $\Delta$ defined by the set $$\Delta=\lbrace (x_1 ,...,x_N )\in (\Re^d )^N / x_i =x_j \rbrace$$ such that $x_i =x_j$ for at least one pair.  Here we can imagine that there is a hard core interaction between particles keeping them apart. Then we identify the elements of the configuration space $(x_1 ,...,x_N )$ and $(x_{\pi(1)} ,...,x_{\pi(N)} )$, for any element $\pi$ of the symmetry group $S_N$ since particles are identical.
In the special case two-dimensional space with two identical particles, the configuration space $M^2 _2$ 
$$
M^2 _2 =\Re^2 \times\lbrace \mbox{cone without the tip} \rbrace 
$$
is infinitely connected. It is constructed by replacing the coordinates $x_1$ and $x_2$ by the center of mass coordinate $X=\frac{x_1 +x_2 }{2}$ and the relative coordinate $x=x_1 - x_2$ and by removing the diagonal $x_1 = x_2$ means leaving out the origin of the $x$-plane and "modding" by $S_2$ means identifying $x$ and $-x$. Then, the resulting construction is the surface of a cone with the tip $x=0$ excluded. Consequently, any closed loop on the mantle of the cone encircling the tip can not be shrunk to a point. Thus  $M^2 _2$ is multiply connected.
Secondly, the connection to statistics comes through recognizing that the class of closed loops $\sigma_i$ corresponds to an interchange of particles $i$ and $i+1$ (Figure 1). In $d\ge 3$, these loops can be deformed into each other; e.g. by rotating the loop around a diameter of a sphere, then $\sigma_i =\sigma_i ^{-1}$. In $d=2$, this can be done in two homotopically inequivalent ways which can be represented by the loop $C_i C_{i+1}$ where the two particles move either counterclockwise (corresponding to $\sigma_i $) or clockwise (corresponding to $\sigma_i ^{-1}$) interchanging their places, so $\sigma_i \ne\sigma_i ^{-1}$ and they are elements of the braid group $B_N$. The latter condition is the difference between $B_N$ and $S_N$.

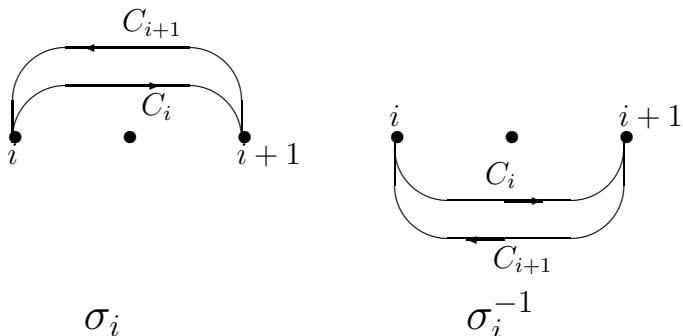
\begin{center}

\setlength{\unitlength}{.2in}
\begin{picture}(20,6)

\put(3.13,0.07){\oval(6,5)[t]}

\put(3.13,0.07){\oval(6,3)[t]}

\put(3,2.55){\vector(-1,0){1}}

\put(3,1.55){\vector(1,0){1}}

\put(3,0){$\bullet$}

\put(6,0){$\bullet$}

\put(0,0){$\bullet$}

\put(0,-0.5){$i$}

\put(6,-0.5){$i+1$}

\put(3.5,.8){$C_{i}$}

\put(3,3){$C_{i+1}$}

\put(2,-4.8){\Large{$\sigma_i$} }

\put(13.13,0.07){\oval(6,5)[b]}

\put(13.13,0.07){\oval(6,3)[b]}

\put(13,-2.47){\vector(-1,0){1}}

\put(13,-1.47){\vector(1,0){1}}

\put(13,0){$\bullet$}

\put(16,0){$\bullet$}

\put(10,0){$\bullet$}

\put(10,0.5){$i$}

\put(16,0.5){$i+1$}

\put(12.5,-1){$C_i $}

\put(12.7,-3.2){$C_{i+1}$}

\put(12,-4.8){\Large{$\sigma_{i}^{-1}$}}

\end{picture}

\end{center}
\vskip 3cm
\hspace*{1cm}Figure 1: The interchange of two particles $i$ and $i+1$ along the closed loop $C_{i}C_{i+1}$.\\
\vskip 1cm

Now to look for a unitary one-dimensional representations of $B_N$, we pose $\chi (\sigma_i)=e^{i\phi_i }$ and we have the following constraint
$$\sigma_i \sigma_{i+1}\sigma_i =\sigma_{i+1}\sigma_i \sigma_{i+1}$$
from the Feynman propagator $$K=\sum\limits_{\al\in B_N}\chi(\al)K^\al $$ requiring $\chi (\sigma_i)\chi (\sigma_j)=\chi (\sigma_i\sigma_j)$. The representations $\chi(\al)$ are the weights of different classes $\al$ and the sum runs over all classes. $K^\al$ denotes the integral over all paths in the class $\al$. The constraint $
\sigma_i \sigma_{i+1}\sigma_i = \sigma_{i+1}\sigma_i \sigma_{i+1}$ requires that $\phi_i =\phi_{i+1}$. Thus it is customary to write
\beq
\chi (\sigma_i)=e^{-i\nu\pi },\phantom{~~~}\chi (\sigma_i ^{-1})=e^{i\nu\pi },\phantom{~~~~~~~~~~~~~~~}\nu\in\lbrack 0,2)
\eeq
with $\nu$ is called statistical parameter. If $\nu=0$, the particles are bosons that obey Bose-Einstein statistics and if $\nu=1$ the particles are fermions with Fermi-Dirac statistics.
After reviewing in brief the planar system and its statistics, we give in what follows its associated symmetry interpolating between bosonic and fermionic symmetries.

\section{Two-Dimensional Particles Algebra}
\hspace{.3in}In this section, we start by briefly recalling the non-commutative geometry \cite{c}. Its most simple example consists of the geometric space described by non-commutative hermitian operator coordinates $x_{i}$, and by considering the non-commutative momentum operators $p_i=i\partial_{x_i}$ $(\partial_{x_i}$ the corresponding derivative of $x_i$). These operators satisfy the following algebra
\beq
\bra{cc}
\lbrack x_{i} ,x_{j} \rbrack = i\theta\epsilon_{ij},\phantom{~~~~~}
\lbrack p_i,p_j \rbrack =i\theta^{-1}\epsilon_{ij},\phantom{~~~~~}\lbrack p_i,x_j \rbrack =-i\delta_{ij}\\\\
\lbrack p_i ,t\rbrack =0=\lbrack x_i ,t\rbrack ,\phantom{~~~~~}\lbrack p_i,\partial_{t}\rbrack =0=\lbrack x_i,\partial_{t}\rbrack,
\era
\eeq
with $t$ the physical time and $\partial_{t}$ its corresponding derivative.

By considering two-dimensional harmonic oscillator which can be decomposed into one-dimensional oscillators. So, it is known that the algebra (2) allows to define, for each dimension, the representation of annihilation and  creation operators as follows
\beq
\bra{ll}
a_{i}&=\sqrt{\frac{\mu\om}{2}}(x_i +\frac{i}{\mu\om} p_i ) \\
a_{i}^{\da}&=\sqrt{\frac{\mu\om}{2}}(x_i -\frac{i}{\mu\om}  p_i ) .\\
\era
\eeq
with $\mu$ is the mass and $\om$ the frequency. These operators satisfy $$\lbrack a_i ,a^{\da}_i \rbrack =1,$$ defining the Heisenberg algebra. In the simultanuously non-commutative space-space and non-commutative momentum-momentum, the bosonic statistics should be maintained; i.e, the operators $a_{i}^{\da}$ and $a_{j}^{\da}$ are commuting for $i\ne j$. Thus, the deformation parameter $\te$ is required to satisfy the condition $$\te=-(\frac{1}{\mu\om})^2\te^{-1}.$$

To find out an algebra describing the planar system we start by introducing the non-commutative geometry depending on the statistical parameter $\nu\in\bf R\rm$. In this investigation we don't deform the combined commutator of momentum and spatial coordinates. The noncommutative geometry is then defined by the following fundamental algebra\\

{\bf Prposition 1}
\beq
\bra{cc}
\lbrack x_i ,x_j \rbrack_\chi = i\theta \epsilon_{ij},\phantom{~~~~~} \lbrack p_i ,p_j \rbrack_\chi =- i\theta (\mu\om)^{2}\epsilon_{ij},\phantom{~~~~~}\lbrack p_i ,x_j \rbrack =-i\delta_{ij} \\\\
\lbrack p_i ,t\rbrack =0=\lbrack x_i ,t\rbrack ,\phantom{~~~~~}\lbrack p_i,\partial_{t}\rbrack =0=\lbrack x_i,\partial_{t}\rbrack.
\era
\eeq
By straightforward calculations we obtain
\beq
\bra{lr}
\lbrack x_i ,p_j\rbrack_\chi =i\de_{ij}+C_{ji},&
\lbrack p_i ,x_j \rbrack_\chi =-i\de_{ij}+D_{ji},
\era
\eeq
where $C_{ji}=(1-\chi)p_j x_i$ and $D_{ji}=(1-\chi)x_j p_i$. The second deformation parameter $\chi$ is given by\\

{\bf Definition 1}
\beq
\chi=e^{\pm i\nu\pi},
\eeq
where $\pm$ sign indicates the two rotation directions on two-dimensional space. $\te$ is non-commutative parameter depending on statistical parameter $\nu$ as we will see later and the notation $[x,y]_q =xy-qyx$.

Then, we introduce an operator $\xi_i$ acting on the momentum direction in the phase-space. We assume that $\xi_i$ satisfies the following commutation relation\\

{\bf Proposition 2}
\beq
\lbrack\xi_i ,x_{j}\rbrack =0\phantom{~~~~~}\forall i,j.
\eeq
In this case, we define the annihilation and the creation operators by\\

{\bf Definition 2}
\beq
\bra{ll}
b_{i}^-&= \sqrt{\frac{\mu\om}{2}}(x_i +\frac{i}{\mu\om} \xi_i p_i ) \\ \\
b^+_{i}&= \sqrt{\frac{\mu\om}{2}}(x_i -\frac{i}{\mu\om} \xi^{-1}_i p_i ) ,
\era
\eeq
with $\xi_i $ is defined in terms of statistical parameter $\nu$ and an operaor $K_i $ which could be a function of the number operator $N$ \\

{\bf Definition 3}
\beq
\xi_i =e^{i\nu\pi K_i },
\eeq
To be consistent with the hermiticty of coordinates we suggest that $\te$ is satisfying $$\theta^\dagger=\chi^{-1}\theta.$$ Then, we give the following expression\\

{\bf Definition 4}

\beq
\te =\nu(1+\chi)I.
\eeq

Consequently the exotic particles algebra is defined by the following commutation relations
\beq
\bra{lll}
\lbrack b^-_{i},b^+_{j} \rbrack_\chi &= \frac{1}{2}(\xi_i +\xi^{-1}_j ) \de_{ij}+i\frac{\mu\om}{2}\te(I+\xi_i \xi^{-1}_j)\epsilon_{ij} -\frac{i}{2}(\xi_j^{-1}C_{ji}-\xi_i D_{ji}),\\ \\
\lbrack b^+_{i},b^+_{j} \rbrack_\chi &= \frac{1}{2}(\xi^{-1}_j -\xi^{-1}_i ) \de_{ij}+ i\frac{\mu\om}{2}\te(I-\xi^{-1}_i \xi^{-1}_j ) \epsilon_{ij} -\frac{i}{2}(\xi_j^{-1}C_{ji}+\xi_i^{-1}D_{ji}),\\ \\
\lbrack b^-_{i},b^-_{j} \rbrack_\chi &= \frac{1}{2}(\xi_i -\xi_j ) \de_{ij}+i\frac{\mu\om}{2}\te(I-\xi_i \xi_j )\epsilon_{ij}+\frac{i}{2}(\xi_j C_{ji}+\xi_i D_{ji}).
\era
\eeq
It is clear that this algebra is a deformed version of Heisenberg algebra and this is one of the very interesting results discussed in this work.
Another important remark we get from this symmetry is concerning the values that statistical parameter can take leading to extremes; if $\nu=0$ we get $\chi=1$, $\te=0$ and $C_{ji}=0=D_{ji}$ and the symmetry (11) becomes bosonic algebra generated by the operators given by (3). Then, this latter is an extreme of the symmetry (11) describing exotic particles system. Another interesting case is $\nu=1$ leading to $\chi=-1$, $\te=0$ and $C_{ji}\ne 0\ne D_{ji}$ and the algebra (11) goes a deformed fermionic algebra defined by
\beq
\bra{ccc}
\lbrace b^-_{i},b^+_{j} \rbrace = \frac{1}{2}(e^{i\pi K_i} +e^{-i\pi K_j} ) \de_{ij} -\frac{i}{2}(e^{-i\pi K_j}C_{ji}-e^{i\pi K_i} D_{ji}),\\ \\
\lbrace b^+_{i},b^+_{j} \rbrace = \frac{1}{2}(e^{-i\pi K_j} -e^{-i\pi K_i} ) \de_{ij}-\frac{i}{2}(e^{-i\pi K_j}C_{ji}+e^{-i\pi K_i}D_{ji}),\\ \\
\lbrace b^-_{i},b^-_{j} \rbrace = \frac{1}{2}(e^{i\pi K_i} -e^{i\pi K_j} ) \de_{ij}+\frac{i}{2}(e^{i\pi K_j} C_{ji}+e^{i\pi K_i} D_{ji}).
\era
\eeq
as a second extreme of (11).

The main results we get from this section is that exotic particles algebra goes to bosonic algebra if $\nu\longrightarrow 0$. This means that our system is originally gotten by exciting a bosonic system in two-dimensional space. Also, we get a deformed fermionic algebra as a second extreme when the statistical parameter $\nu$ equals to 1. Thus, we remark that the system described by the above algebras (12) doesn't have anything to do with fermions originally but it could be related to something else as deformed fermions which are known in the literature as quionic particles or $k_i$-fermions, $k_i$ integer number introduced as deformation parameter, and these kinds of particles are known as non physical particles.
\section{Deformed Oscillator Algebras}
\hspace*{.3in}In this section we show that the extended Heisenberg algebra \cite{eha} could be a symmetry of planar system at defomed level. First we start by a short review on $C_{\lambda}$-extended oscillator algebra and then we give the defomed form of this symmetry which describes exotic particles in two-dimensional space.
\subsection{Extended Heisenberg Algebra}
\hspace*{.3in}We review in brief the $C_{\lambda}$-extended oscillator algebras. As known in the literature, a generalization of the Calogero-Vasiliev algebras, the $C_{\lambda}$-extended oscillator algebras (also
called generalized deformed oscillator algebras (GDOA's)), denoted $A^{\lambda}$, $\lambda=2,3,...,$ are defined by
\beq \bra{lrlr}
[N,a^{\dagger}]=a^{\dagger},& [a,a^{\dagger}]=I+\sum\limits_{\mu=0}^{\lambda-1}\alpha_{\mu}P_{\mu},\\ \\

[N,P_{\mu}]=0,& a^{\dagger}P_{\mu}=P_{\mu +1}a^{\dagger},
\era \eeq together with their hermitian conjugates, and \beq
\bra{lrlr}
P_{\mu}=\frac{1}{\lambda}\sum\limits_{\nu=0}^{\lambda-1}
e^{\frac{2\pi i\nu(N-\mu)}{\lambda}},&
\sum\limits_{\mu=0}^{\lambda-1}P_{\mu}=1,&
P_{\mu}P_{\nu}=\de_{\mu,\nu}P_{\nu}\\ \\
\sum\limits_{\mu=0}^{\lambda-1}\alpha_{\mu}=0,&
\sum\limits_{\nu=0}^{\mu-1}\alpha_{\nu}>-1,& \mu=1,...,\lambda-1.
\era, \eeq  where $\alpha_{\mu}\in{\bf R\rm}$, $N$ is the number
operator and $P_{\mu}$ are the projection operators on subspaces
$F_{\mu}=\{\vert k\lambda -\mu\rangle \vert k=0,1,2,...\}$ of the
Fock space $F$ which is portioning into $\lambda$ subspaces.

The operators $a$ and $a^{\dagger}$ are defined by \beq \bra{lrlr}
a^{\dagger}a=F(N),& aa^{\dagger}=F(N+1), \era \eeq where
$F(N)=N+\sum\limits_{\mu=0}^{\lambda-1}\be_{\mu}P_{\mu}$,
$\be_{\mu}=\sum\limits_{\nu=0}^{\mu-1}\al_{\nu}$, which is a
fundamental concept of deformed oscillators. Let's  denote the
basis states of subspaces $F_{\mu}$ by $\vert n\rangle=\vert
k\lambda +\mu\rangle\simeq (a^{\dagger})^n\vert 0\rangle$ where
$a\vert 0\rangle=0$, $\vert 0\rangle$ is the vacuum state. The
operators $a$, $a^{\dagger}$  and $N$ act on $F_{\mu}$ as
follows \beq \bra{rcl} N\vert n\rangle=n\vert n\rangle,&
a^{\dagger}\vert n\rangle=\sqrt{F(N+1)}\vert n+1\rangle,& a\vert
n\rangle=\sqrt{F(N)}\vert n-1\rangle. \era \eeq According to these
relations, $a$ and $a^{\dagger}$ are the
annihilation and the creation operators respectively.

Particularly, if $\lambda=2$, we have two projection operators
$P_{0}=\frac{1}{2}(I+(-1)^{N})$ and
$P_{1}=\frac{1}{2}(I-(-1)^{N})$ on the even and odd subspaces of
the Fock space $F$, and the relations of (13) are restricted to
\beq \bra{rcl} [N,a^{\dagger}]=a^{\dagger},&
[a,a^{\dagger}]=I+\kappa K,& \{K,a^{\dagger}\}=0, \era \eeq with
their hermitian conjugates, where $K=(-1)^{N}$ is the Klein
operator and $\kappa$ is a real parameter. These relations define
the so-called Calogero-Vasiliev algebra.

The $C_{\lambda}$-extended oscillator algebras are seeing as
deformation of $G$-extended oscillator algebras, where $G$ is some
finite group, appeared in connection with $n$-particle integrable
models. In the former case, $G$ is the symmetric group $S_n$. So,
for two particles $S_2$, can be realized in terms of $K$ and
$S_2$-extended oscillator algebra becomes a generalized
deformed oscillator algebra (GDOA) also known as the
Calogero-Vasiliev or modified oscillator algebra. In the
$C_{\lambda}$-extended oscillator algebras, $G\equiv C_{\lambda}$
is the cyclic group of order $\lambda$,
$C_{\lambda}=\{1,K,...,K^{\lambda-1}\}$. So, these algebras have a
rich structure since they depend upon $\lambda$ independent real
parameters, $\alpha_0 ,\alpha_1
,...,\alpha_{\lambda-1},$.
\subsection{Deformed $C_{\lambda}$-Extended Heisenberg Algebra}
\hspace*{.3in}Other version of anyonic algebra can be obtained by treating the special case of statistical parameter $\nu\in[0,1]$. By using, the Taylor expansion to the operator $\xi_i $, the first commutation relation in the algebra (11) can be rewritten in this form
\beq
\lbrack b^-_{i},b^+_{j} \rbrack_\chi =(I+\Re_{ij}^{[\nu]})\de_{ij}+A^{[\nu]}_{ij}\ep_{ij}+Q^{[\nu]}_{ij},
\eeq
where
\beq
\Re_{ij}^{[\nu]}=\sum\limits_{\ell=1}^{\frac{n+1}{2}}\kappa_{\nu,\ell}\frac{K_i ^{2\ell -1}-K_j ^{2\ell -1}}{2}+\sum\limits_{k=1}^{\frac{m}{2}}\kappa_{\nu,k}\frac{K_i ^{2k }+K_j ^{2k}}{2},
\eeq
$$Q^{[\nu]}_{ij}=-\frac{i}{2}\sum\limits_{p=0}^{\lambda -1}\frac{(i\nu\pi)^p }{p!}((-K_j)^{p}C_{ji}-K_i^p D_{ji})$$
and 
\beq
A^{[\nu]}_{ij}=\frac{i\te\mu w}{2}\Big( I+\sum\limits_{\al=0}^{\lambda -1}\frac{(i\nu\pi)^\al }{\al!}(K_i -K_j )^\al\Big) ,
\eeq
with $m\in\bf N \rm$ is even and $n\in\bf N \rm$ odd such that $n,m\le\lambda -1$ with $\lambda\in\bf N \rm$ by imposing $$K_i ^{\lambda} = I.$$ 
 The coeffecients $\kappa_{\nu,\ell}$ and $\kappa_{\nu,k}$ are given in terms of statistical parameter as follows
$$\kappa_{\nu,\ell}=\frac{(i\nu\pi)^{2\ell -1}}{(2\ell -1)!},\phantom{~~~~}\kappa_{\nu,k}=\frac{(i\nu\pi)^{2k}}{(2k)!}.$$
Then, the last two commutation relations of (11) become
\beq
\bra{ll}
\lbrack b^+_{i},b^+_{j} \rbrack_\chi &=  \sum\limits_{\al=1}^{\lambda -1}\frac{(-i\nu\pi)^\al }{\al!}\frac{K_j^\al -K_i^\al }{2} \de_{ij}-i\frac{\mu\om}{2}\te\sum\limits_{\al=1}^{\lambda -1}\frac{(-i\nu\pi)^\al }{\al!}(K_i +K_j )^\al \epsilon_{ij}\\\\&-
\frac{i}{2}\sum\limits_{p=0}^{\lambda -1}\frac{(-i\nu\pi)^p }{p!}(K_j^{p}C_{ji}-K_i^p D_{ji}),\\ \\
\lbrack b^-_{i},b^-_{j} \rbrack_\chi &=  \sum\limits_{\al=1}^{\lambda -1}\frac{(i\nu\pi)^\al }{\al!}\frac{K_i^\al -K_j^\al }{2} \de_{ij}-i\frac{\mu\om}{2}\te\sum\limits_{\al=1}^{\lambda -1}\frac{(i\nu\pi)^\al }{\al!}(K_i +K_j )^\al \epsilon_{ij}\\\\&+\frac{i}{2}\sum\limits_{p=0}^{\lambda -1}\frac{(i\nu\pi)^p }{p!}(K_j^{p}C_{ji}-K_i^p D_{ji})
.
\era
\eeq
Now if we pose the following\\

{\bf Proposition 3}
$$K_i =e^{i\frac{2\pi}{\lambda}N_i }$$
with $N_i $ is a number operator defined in terms of $b^+_{i}$ and $b^-_{i}$ as
\beq
b^+_{i}b^-_{i}=f(N),\eeq
with $f$ is some function such that the following relations are satisfied
$$\bra{ll}
\lbrack N_i, b^-_{j}\rbrack = -\de_{ij}b_{i},\phantom{~~~~~~}\lbrack N_i, b^+_{j}\rbrack = \de_{ij}b^{\da}_{i},\\\\
K_i b_{j} = \de_{ij}e^{-i\frac{2\pi}{\lambda}}b_{j}K_i ,\phantom{~~~~~~}K_i b^{\da}_{j} = \de_{ij}e^{i\frac{2\pi}{\lambda}}b^{\da}_{j}K_i.
\era
$$

Thus, It is easy to see that the obtained relations (18) and (21) define a physicswise realization of deformed $C_{\lambda}$-extended Heisenberg algebra on two-dimensional non-commutative space describing exotic particles, where $C_{\lambda}$ is a cyclic group $$C_{\lambda}=\{I, K_i , K^2_i ,K^3_i ,...,K^{\lambda-1}_i \},\phantom{~~~~}\lambda\in\bf N\rm.$$ Again once, it is clear that the obtained algebra goes to bosonic symmetry if $\nu$ goes to 0.
\subsection{Fock Representation}
\hspace{.3in}It is convenient to construct a Fock representation for the algebra (4,5) underlying the non-commutative geometry by way of the operators (8) obeying (18,21). The Fock space is introduced by the set
\beq
F_i=\{\mid n\rangle;\phantom{~~}n=0,1\}
\eeq
with the states $\mid n\rangle$ are defined as
\beq
\mid n\rangle=\frac{1}{f(\sqrt{n!})}(b_i^+ )^n \mid 0\rangle,\phantom{~~~~}n=0,1
\eeq
they are the quantum mechanical states inherent to the non-commutativity (4), with $\mid 0\rangle$ the vacuum state and $f$ is some funtion definning the number operator $N_i$ as given above in (22). The "exotic" annihilation and creation operators act on Fock space as
\beq\bra{ll}
b^+_{i}\mid n\rangle =f(\sqrt{n+1}) \mid n+1\rangle,\\\\
b^-_{i}\mid n\rangle =f(\sqrt{n}) \mid n-1\rangle.
\era\eeq
Then, Owing to (8,25), the non-commuting spatial and momentum coordinates are acting on Fock space as
\beq\bra{ll}
x_i \mid n\rangle =\frac{1}{\sqrt{2\mu w}}\Big[ f(\sqrt{n}) \mid n-1\rangle+ f(\sqrt{n+1}) \mid n+1\rangle \Big],\\\\
p_i \mid n\rangle =\frac{i\sqrt{2\mu w}}{e^{i\frac{2\pi}{\lambda}n }-e^{-i\frac{2\pi}{\lambda}n }}\Big[ -f(\sqrt{n}) \mid n-1\rangle+ f(\sqrt{n+1}) \mid n+1\rangle \Big],

\era\eeq

\section{$\om_\infty$-Algebra}
\hspace{.3in}First let us extend the commutation relation of (18). For any order $\be$ of $b^{\da}_j$ we obtain the following extended commutation relation
on non-commutative space
\beq
\lbrack b_{i},(b^{\da}_{i})^{\be} \rbrack_{\chi} = \Big ( \be +F_{ii}+P_{ii}\Big )  (b^{\da}_{i})^{\be -1},
\eeq
where
$$
F_{ii}=\sum\limits_{k=1}^{\frac{m}{2}}p_{k}\kappa_{\nu,k}K_i ^{2k },\phantom{~~~~~~}P_{ii}=-\frac{i}{2}\sum\limits_{p=0}^{\lambda -1}\frac{(i\nu\pi)^p }{p!}p_{k}K_i^p((-1)^{p}C_{ii}- D_{ii})
$$
with $p_{k}=\sum\limits_{\ga=0}^{\be-1}e^{\frac{-2(2k)\pi\ga}{\lambda}}$.\\

Then we generalize the symmetry (27) for any order $\al$ and $\be$ of $b_i$ and $b^{\da}_i$ respectively and we get
\beq
\lbrack b_i ^{\al},(b^{\da}_{i})^{\be}\rbrack_{\chi} = \sum\limits^{\al-1}_{k=0}\sum\limits^{k}_{\eta=0}\Big ( \be +(F_{ii}+P_{ii})q^{k}\Big ) G_{ii}^{[\eta]}(b^{\da}_{i})^{\be-\eta -1}(b_{i})^{\al-\eta -1},
\eeq
with $q=e^{\frac{-i2\pi}{\lambda}}$ and according to (27) the operators $G_{ii}^{[\eta]}$ are defined by the following relation
\beq\bra{ll}
(b_{i})^{\al}(b^{\da}_{i})^{\be-1}&=G_{ii}^{[\al]}(b^{\da}_{i})^{\be-\al-1}+G_{ii}^{[\al-1]}(b^{\da}_{i})^{\be-\al}b_{i}+...\\ \\
&+G_{ii}^{[0]}(b^{\da}_{i})^{\be-1}(b_{i})^{\al},
\era\eeq
here we recite some of them, since the calculations are long and complicate for the orders $9<\al<\al-5$,
\begin{displaymath}
\bra{llllll}
G_{ii}^{[0]}&=1\\
G_{ii}^{[1]}&=\al(\be-1)+\sum\limits_{\epsilon=0}^{\al-1}(F_{ii}+P_{ii})q^{\epsilon}\\
&.\\
&.\\
&.\\
G_{ii}^{[\al-1]}&=\prod\limits_{\sigma=1}^{\al-1}\Big ( (\be-\sigma)+(F_{ii}+P_{ii})q^{\al-\si}\Big ) +\\ \\
&+\sum\limits^{\al-2}_{\mu=2}\prod\limits_{\sigma=1}^{\al-\mu}\Big ( (\be-\sigma)+(F_{ii}+P_{ii})q^{\al-\si}\Big ) \prod\limits_{\sigma=\al-(\mu-1)}^{\al-1}\Big ( (\be-\sigma)+(F_{ii}+P_{ii})q^{\mu-\al-1}\Big ) \\ \\
&+\Big ( 2(\be-1)+\sum\limits^{\al-1}_{\eta=\al-2}(F_{ii}+P_{ii})q^{\eta}\Big ) \prod\limits_{\sigma=2}^{\al-1}\Big ( (\be-\sigma)+(F_{ii}+P_{ii})q^{\al-\si-1}\Big ) \\ \\
G_{ii}^{[\al]}&=\prod\limits_{\sigma=1}^{\al}\Big ( (\be-\sigma)+(F_{ii}+P_{ii})q^{\al-\si}\Big ).
\era
\end{displaymath}

Thus, in terms of quasi-particles operators (8), we define the following generators
\beq
T_{i,m} ^s =(b^{\da}_{i})^s (b_{i})^m .
\eeq

By using the commutation relations (28) and the definition (29) of the $G_{ij}^{[\eta]}$, the operatos $T_{i,m} ^s $ satisfy the commutation relaions
\beq
\lbrack T_{i,m} ^s , T_{i,n} ^t \rbrack_{\chi}=\Big [ \sum\limits^{m-1}_{\al=0}\sum\limits^{\al}_{\ell=0} R^{t,s}_{\al,\ell}- \sum\limits^{n-1}_{\al=0}\sum\limits^{\al}_{\ell=0}R^{s,t}_{\al,\ell} \Big ] T_{i,m+n-(\ell-1)} ^{s+t-(\ell-1)},
\eeq
with $$R^{t,s}_{\al,\ell}=\Big ( t +(F_{ii}+P_{ii})q^{\al+s}\Big ) G_{ii(s)}^{[\ell]},$$ where the notation $G_{ii(s)}^{[\ell]}$ indicates that the parameter $q$ in the expression of $G_{ii}^{[\ell]}$ becomes $q^s$.\\

The relation (31) defines a $\om_\infty$ algebra characterizing the non-commutative system consisting of two-dimensional particles.\\
\section{Conclusion}
\hspace{.3in}We conclude by giving the main results of this paper: The exotic particles algebra is constructed basing on noncommutative geometry depending on statistical parameter which characterizes the planar system. In this background, the deformed version of $C_\lambda$-extended Heisenberg algebra is realized in physicswise having two extremes bosonic algebra and deformed fermionic algebra. Then using the "exotic" annihilation and creation operators generating the planar system algebra we construct a $w_\infty$ symmtery characterizing the two-dimensional noncommutative system.\\

Acknowledgements: The author would like to thank the Abdus Salam ICTP for the hospitality.

\end{document}